\documentclass[a4paper,aps,prd,10pt,preprintnumbers,showpacs,twocolumn,superscriptaddress,nofootinbib,amsmath,amssymb]{revtex4-1}
\usepackage[dvips]{graphics}
\usepackage{cmap}
\usepackage[utf8]{inputenc}
\usepackage[T1]{fontenc}

\def\imo{i}

\def\K{{\cal K}}
\def\Order#1{{\cal O}\left(#1\right)}

\begin{document}
\title{Quasinormal modes of charged black holes in Asymptotically Safe Gravity}
\author{Alexey Dubinsky}\email{dubinsky@ukr.net}
\affiliation{University of Sevilla, 41009 Seville, Spain}
\begin{abstract}
We calculate quasinormal modes of scalar and neutrino perturbations around the charged black hole in Asymptotically Safe Gravity. We show that the charge and coupling constant change the quasinormal spectrum considerably. We show that previous calculations of scalar quasinormal modes in this background [F. Javed, Phys. Dark Univ., 44, 101450 (2024)] suffer from a large numerical error exceeding the effect, that is, the deviations of the frequencies from their Schwarzschild limits. In the high frequency (eikonal) limit an explicit analytic formula for quasinormal modes is derived, which confirms the correspondence between the null circular geodesics and eikonal quasinormal frequencies. 
\end{abstract}
\maketitle
\section{Introduction}

Asymptotically safe gravity \cite{Niedermaier:2006wt,Percacci:2007sz} is a theoretical framework within the field of quantum gravity that posits the existence of a non-trivial ultraviolet (UV) fixed point in the renormalization group flow of the theory. In simpler terms, it suggests that the theory remains well-defined and consistent at all energy scales, including extremely high energies where quantum effects become significant.

In conventional quantum field theories, such as quantum electrodynamics, infinities arise in calculations when attempting to describe particle interactions at very high energies. These infinities indicate that the theory breaks down and cannot be used to make meaningful predictions beyond a certain energy scale.

However, in asymptotically safe gravity, it is proposed that the theory remains valid even at these extremely high energies. This is achieved by introducing additional terms in the theory's action that contribute to the renormalization group flow in such a way that the theory approaches a stable fixed point at high energies. At this fixed point, the theory becomes "asymptotically safe," meaning that it remains well-defined and predictive even at energies where quantum effects are significant.

Probably the first black hole model within the asymptotically safe gravity was suggested by Bonanno and Reuter in \cite{Bonanno:2000ep}, where the main corrections come from the gravitational constant which becomes dependent upon the distance. This black hole metric, initially formulated for an electrically neutral black hole was generalized to the case of non-zero electric charge $Q$ \cite{Ruiz:2021qfp,Ladino:2023zdn}. 
One of the most interesting feature of these black hole models is absence of central singularity, which takes place for ordinary Schwarzschild solution.

At the same time, the basic characteristic of a black is its spectrum of damped oscillations, called {\it quasinormal modes} \cite{Kokkotas:1999bd,Berti:2009kk,Konoplya:2011qq}, 
they are observed in current experiments with gravitational interferometers \cite{LIGOScientific:2016aoc,LIGOScientific:2017vwq,LIGOScientific:2020zkf} and still large uncertainty in our knowledge of the angular momentum and mass of the resultant ringing black hole leaves a big room for alternative theories of gravity.
Consequently, perturbations and quasinormal modes of various regular black holes have been studied in a great number of recent papers \cite{Bronnikov:2019sbx,Mukohyama:2023xyf,Konoplya:2023ppx,Rayimbaev:2022mrk,DuttaRoy:2022ytr,Al-Badawi:2023lke,Pedraza:2021hzw,Lin:2013ofa,Bolokhov:2023ruj,Pedrotti:2024znu,MahdavianYekta:2019pol,Hendi:2020knv,Filho:2023voz,Li:2016oif,Toshmatov:2015wga,Jawad:2020hju,Macedo:2016yyo,Zhang:2024nny,Wahlang:2017zvk,Rincon:2020cos,Lopez:2022uie,Gingrich:2024tuf}.
Special attention has been devoted to spectra of black holes in the Asymptotically Safe Gravity \cite{Rincon:2020iwy,Konoplya:2022hll,Konoplya:2023aph,Lambiase:2023hng,Zinhailo:2023xdz,Malik:2024tuf}.

The quasinormal modes of the charged black hole in the Asymptotically Safe Gravity have been recently calculated with the help of the WKB formula in the eikonal regime \cite{Javed:2024hzs}. In the present research we will show that the results in \cite{Javed:2024hzs} contain a large numerical error, not allowing to properly estimate the dominant quasinormal frequencies of the scalar field. On the countrary, we perform comprehensive calcuations of scalar and neutrino quasinormal modes with the help of two independent and sufficiently accurate methods: the 6th order WKB method with Padé approximants and the time-domain integration with consequent Prony extraction of frequencies from the time-domain profiles. The results from both methods are in close agreement. The quasinormal modes are significantly influenced by both the coupling constant and charge.
In addition we derive the explicit analytic formula in the eikonal limit and show that it respects the correspondence between the null geodesics and eikonal quasinormal modes.
While the correspondence was mentioned in \cite{Javed:2024hzs}, and no explicit expression for quasinormal modes in terms of the black hole parameters were presented there.

The paper is organized as follows. In sec.  \ref{sec:wavelike} we summarize the main information about the metric, the wave-like euqations and effective potentials, Sec. \ref{sec:methods} reviews the methods used for calculations of quasinormal modes: time-domain integration method and WKB approach. Sec. \ref{sec:QNM} is devoted to calculations of quasinormal modes and in sec.  \ref{sec:conclusions} we summarize the obtained results.

\section{Black hole metric, wave equations and effective potentials}\label{sec:wavelike}

The line element describing the metric of the quantum-corrected charged black hole is expressed as \cite{Ruiz:2021qfp,Ladino:2023zdn}:
\begin{equation}\label{metric}
  ds^2=-f(r)dt^2+\frac{dr^2}{f(r)}+r^2(d\theta^2+\sin^2\theta d\phi^2),
\end{equation}
where
$$
\begin{array}{rcl}
f(r)&=&\displaystyle 1 -\frac{2 M r}{\gamma +r^2}+\frac{Q^2}{\gamma +r^2},\\
\end{array}
$$
Here, $\gamma $ represents the quantum parameter, and $M$ denotes the ADM mass. All dimensional quantities will be measured in units of mass, where $M=1$ is chosen as the reference.
As the charge approaches zero, the above metric simplifies to the well-known Bonanno-Reuter black hole \cite{Bonanno:2000ep}.

The equations governing the scalar ($\Phi$) and Dirac ($\Upsilon$) fields within the framework of general relativity can be expressed as:
\begin{subequations}\label{coveqs}
\begin{eqnarray}\label{KGg}
\frac{1}{\sqrt{-g}}\partial_\mu \left(\sqrt{-g}g^{\mu \nu}\partial_\nu\Phi\right)&=&0,
\\\label{covdirac}
\gamma^{\alpha} \left( \frac{\partial}{\partial x^{\alpha}} - \Gamma_{\alpha} \right) \Upsilon&=&0,
\end{eqnarray}
\end{subequations}
Here, $F_{\mu\nu}=\partial_\mu A_\nu-\partial_\nu A_\mu$ represents the electromagnetic tensor, $\gamma^{\alpha}$ denotes noncommutative gamma matrices, and $\Gamma_{\alpha}$ signifies spin connections within the tetrad formalism.
Upon separation of variables in the background described by (\ref{metric}), equations (\ref{coveqs}) assume the form of Schrödinger-like equations:
\begin{equation}\label{wave-equation}
\dfrac{d^2 \Psi}{dr_*^2}+(\omega^2-V(r))\Psi=0,
\end{equation}
Here, the "tortoise coordinate" $r_*$ is defined as:
\begin{equation}\label{tortoise}
dr_*\equiv\frac{dr}{f(r)}.
\end{equation}

The effective potentials governing the scalar field ($s=0$) take the form:
\begin{equation}\label{potentialScalar}
V(r)=f(r)\frac{\ell(\ell+1)}{r^2}+\frac{1}{r}\cdot\frac{d^2 r}{dr_*^2},
\end{equation}
Here, $\ell=s, s+1, s+2, \ldots$ represent the multipole numbers.
For the Dirac field ($s=1/2$), there exist two isospectral potentials given by:
\begin{equation}
V_{\pm}(r) = W^2\pm\frac{dW}{dr_*}, \quad W\equiv \left(\ell+\frac{1}{2}\right)\frac{\sqrt{f(r)}}{r}.
\end{equation}
These isospectral wave functions can be transformed into each other via the Darboux transformation:
\begin{equation}\label{psi}
\Psi_{+}\propto \left(W+\dfrac{d}{dr_*}\right) \Psi_{-},
\end{equation}
Thus, it suffices to compute quasinormal modes for only one of the effective potentials. We choose to do so for $V_{+}(r)$ as the WKB method yields better results in this case.

\begin{figure}
\resizebox{\linewidth}{!}{\includegraphics{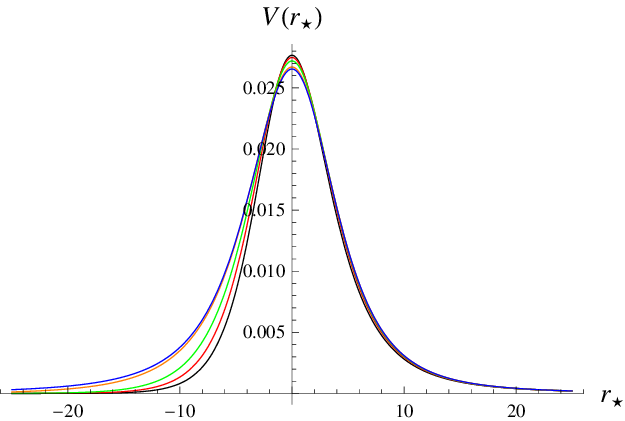}}
\caption{Potential as a function of the tortoise coordinate of the $\ell=0$ scalar field for the quantum corrected charged black hole ($M=1$, $Q=1/2$): $\gamma=0$ (black), $\gamma=0.3$ (red), $\gamma=0.4$ (green), $\gamma=0.7$ (orange), $\gamma=0.75$ (blue).}\label{fig:potentials}
\end{figure}

\begin{figure}
\resizebox{\linewidth}{!}{\includegraphics{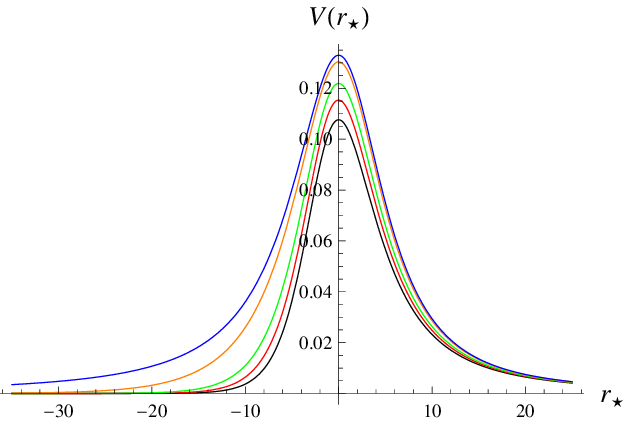}}
\caption{Potential as a function of the tortoise coordinate of the $\ell=1$ scalar field for the quantum corrected charged black hole ($M=1$, $Q=1/2$): $\gamma=0$ (black), $\gamma=0.3$ (red), $\gamma=0.4$ (green), $\gamma=0.7$ (orange), $\gamma=0.75$ (blue).}\label{fig:potentials2}
\end{figure}

\begin{figure}
\resizebox{\linewidth}{!}{\includegraphics{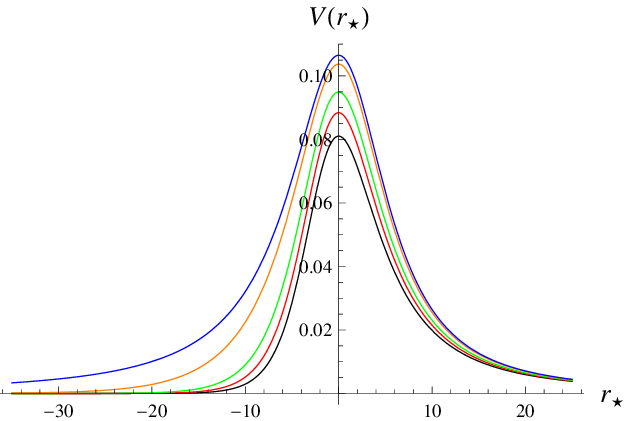}}
\caption{Potential as a function of the tortoise coordinate of the $\ell=1$ scalar field for the quantum corrected charged black hole ($M=1$, $Q=1/2$): $\gamma=0$ (black), $\gamma=0.3$ (red), $\gamma=0.4$ (green), $\gamma=0.7$ (orange), $\gamma=0.75$ (blue).}\label{fig:potentialsEM}
\end{figure}

\begin{figure}
\resizebox{\linewidth}{!}{\includegraphics{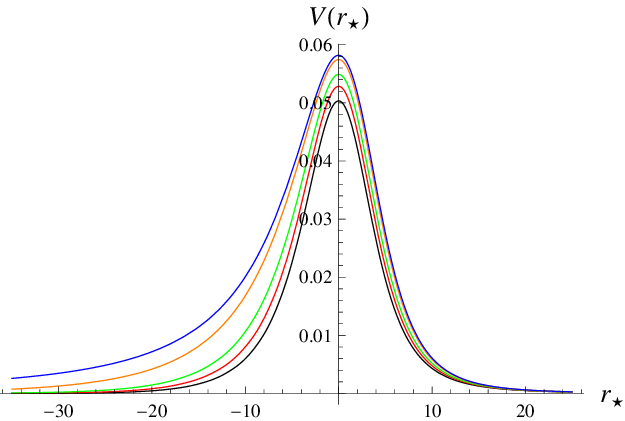}}
\caption{Potential as a function of the tortoise coordinate of the $\ell=1/2$ scalar field for the quantum corrected charged black hole ($M=1$, $\Lambda=1/10$): $\gamma=0$ (black), $\gamma=0.3$ (red), $\gamma=0.4$ (green), $\gamma=0.7$ (orange), $\gamma=0.75$ (blue).}\label{fig:potentials3}
\end{figure}

The effective potentials for the scalar and Dirac fields are shown in figures \ref{fig:potentials}-\ref{fig:potentials3}. The effective potential are positive definite. However, one the two iso-spectral effective potentials, $V_{-}(r)$, for the Dirac field (not shown and used here) has a negative gap near the event horizon.

\section{The methods}\label{sec:methods}

\subsection{WKB approach}

When the effective potential $V(r)$ in the wave-like equation (\ref{wave-equation}), takes the form of a barrier with a single peak, the WKB formula is suitable for determining the dominant quasinormal modes  satisfying the boundary conditions:
\begin{equation}\label{boundaryconditions}
\Psi(r_*\to\pm\infty)\propto e^{\pm\imo \omega r_*},
\end{equation}
which correspond to purely ingoing waves at the horizon ($r_*\to-\infty$) and purely outgoing waves at spatial infinity ($r_*\to\infty$).

The WKB method relies on matching asymptotic solutions, which fulfill the quasinormal boundary conditions (\ref{boundaryconditions}), with the Taylor expansion around the peak of the potential barrier. The first-order WKB formula, representing the eikonal approximation, becomes exact in the limit $\ell \to \infty$. Subsequently, the general WKB expression for the frequencies can be expanded around the eikonal limit as follows \cite{Konoplya:2019hlu}:
\begin{eqnarray}\label{WKBformula-spherical}
\omega^2&=&V_0+A_2(\K^2)+A_4(\K^2)+A_6(\K^2)+\ldots\\\nonumber&-&\imo \K\sqrt{-2V_2}\left(1+A_3(\K^2)+A_5(\K^2)+A_7(\K^2)\ldots\right),
\end{eqnarray}
where the matching conditions for the quasinormal modes imply that
\begin{equation}
\K=n+\frac{1}{2}, \quad n=0,1,2,\ldots,
\end{equation}
Here, $n$ represents the overtone number, $V_0$ is the value of the effective potential at its maximum, $V_2$ is the value of the second derivative of the potential at this point with respect to the tortoise coordinate, and $A_i$ for $i=2, 3, 4, \ldots$ denotes the $i$-th WKB order correction term beyond the eikonal approximation, dependent on $\K$ and derivatives of the potential at its maximum up to the order $2i$. The explicit form of $A_i$ can be found in \cite{Iyer:1986np} for the second and third WKB order, in \cite{Konoplya:2003ii} for the 4-6th orders, and in \cite{Matyjasek:2017psv} for the 7-13th orders. The WKB approach outlined above for determining quasinormal modes and grey-body factors has been extensively utilized at various orders in numerous studies
(see, for instance, \cite{Konoplya:2006ar,Konoplya:2022hbl,Kokkotas:2010zd,Chen:2019dip,Fu:2022cul,Chen:2023akf,Albuquerque:2023lhm,Fernando:2012yw}).

\subsection{Time-domain integration}

The accuracy of the aforementioned analytical formulas can be verified using two methods. Firstly, by comparing them with the 6th order WKB formula with Padé approximants, and secondly, through a more independent approach, employing time-domain integration. For the time-domain integration, we utilized the Gundlach-Price-Pullin discretization scheme \cite{Gundlach:1993tp}:
\begin{eqnarray}
\Psi\left(N\right)&=&\Psi\left(W\right)+\Psi\left(E\right)-\Psi\left(S\right)\nonumber\\
&&- \Delta^2V\left(S\right)\frac{\Psi\left(W\right)+\Psi\left(E\right)}{4}+{\cal O}\left(\Delta^4\right),\label{Discretization}
\end{eqnarray}
Here, the integration scheme involves the points: $N\equiv\left(u+\Delta,v+\Delta\right)$, $W\equiv\left(u+\Delta,v\right)$, $E\equiv\left(u,v+\Delta\right)$, and $S\equiv\left(u,v\right)$. This method has been employed in numerous studies \cite{Konoplya:2020jgt,Konoplya:2005et,Ishihara:2008re,Churilova:2021tgn,Abdalla:2012si,Konoplya:2020jgt,Aneesh:2018hlp,Varghese:2011ku}, affirming its accuracy.

To extract the frequency values from the time-domain profile, we employ the Prony method, which entails fitting the profile data with a sum of damped exponents:
\begin{equation}\label{damping-exponents}
\Psi(t)\simeq\sum_{i=1}^pC_ie^{-i\omega_i t}.
\end{equation}
We assume that the quasinormal ringing stage begins at $t_0=0$ and ends at $t=Nh$, where $N\geq2p-1$. Consequently, relation (\ref{damping-exponents}) holds true for each point of the profile:
\begin{equation}
x_n\equiv\Psi(nh)=\sum_{j=1}^pC_je^{-i\omega_j nh}=\sum_{j=1}^pC_jz_j^n.
\end{equation}
Subsequently, we determine $z_i$ in terms of the known $x_n$ and calculate the quasinormal frequencies $\omega_i$. Quasinormal modes are typically derived from time-domain profiles when the ring-down stage comprises a sufficient number of oscillations. Notably, the higher the multipole number $\ell$, the longer the ringdown period.

\begin{figure}
\resizebox{\linewidth}{!}{\includegraphics{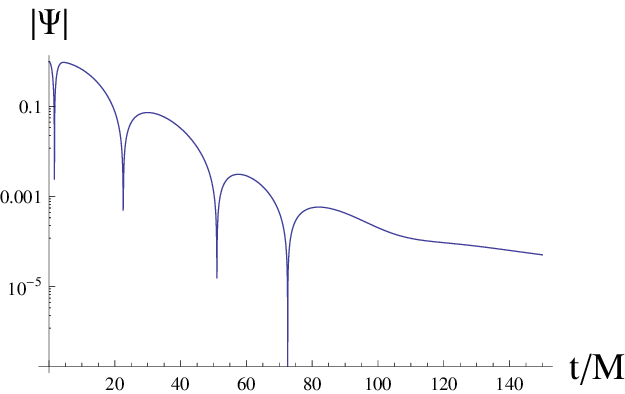}}
\caption{Time-domain profile for the scalar perturbations ($\ell=0$)  $\gamma = 0.1$, $M =1$, $Q/M = 0.5$.}\label{fig:timedomain}
\end{figure}

\begin{figure}
\resizebox{\linewidth}{!}{\includegraphics{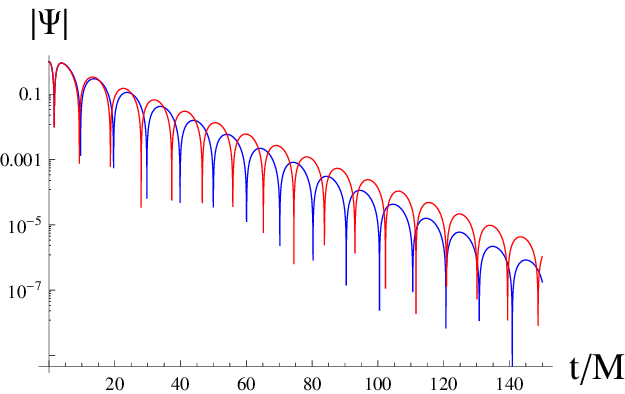}}
\caption{Time-domain profile for the scalar pertubrations ($\ell=1$) $\gamma =0.1 $ (blue) and $\gamma =0.6$ (red); $M =1$, $Q/M = 0.5$.}\label{fig:timedomain2}
\end{figure}

\begin{table*}
\begin{tabular}{c c c c c c}
\hline
$Q$ & $\gamma$ & WKB6 Padé & WKB6 & difference $Re (\omega)$ & difference $Im (\omega)$ \\
\hline
$0$ & $0$ & $0.110792-0.104683 i$ & $0.110467-0.100816 i$ & $0.293\%$ & $3.69\%$\\
$0$ & $0.2$ & $0.114277-0.102400 i$ & $0.114196-0.099072 i$ & $0.0712\%$ & $3.25\%$\\
$0$ & $0.4$ & $0.117619-0.099562 i$ & $0.118397-0.096598 i$ & $0.661\%$ & $2.98\%$\\
$0$ & $0.6$ & $0.120623-0.095727 i$ & $0.124022-0.089448 i$ & $2.82\%$ & $6.56\%$\\
$0$ & $0.7$ & $0.122296-0.092986 i$ & $0.126038-0.084165 i$ & $3.06\%$ & $9.49\%$\\
$0$ & $0.75$ & $0.123063-0.091235 i$ & $0.126160-0.081775 i$ & $2.52\%$ & $10.4\%$\\
$0.5$ & $0$ & $0.116107-0.105495 i$ & $0.115790-0.101961 i$ & $0.273\%$ & $3.35\%$\\
$0.5$ & $0.2$ & $0.120011-0.102341 i$ & $0.120259-0.099867 i$ & $0.207\%$ & $2.42\%$\\
$0.5$ & $0.4$ & $0.123373-0.098171 i$ & $0.126794-0.092256 i$ & $2.77\%$ & $6.03\%$\\
$0.5$ & $0.6$ & $0.126313-0.090288 i$ & $0.127940-0.081224 i$ & $1.29\%$ & $10.0\%$\\
$0.5$ & $0.7$ & $0.124233-0.087455 i$ & $0.125468-0.079889 i$ & $0.995\%$ & $8.65\%$\\
$0.5$ & $0.75$ & $0.123657-0.086991 i$ & $0.125108-0.079391 i$ & $1.17\%$ & $8.74\%$\\
\hline
\end{tabular}
\caption{Quasinormal modes of the $\ell=0$, $n=0$ scalar field for the quantum corrected charged black hole calculated using the 6th WKB method with and without Padé approximants; $M=1$.}\label{table1}
\end{table*}

\begin{table*}
\begin{tabular}{c c c c c c}
\hline
$Q$ & $\gamma$ & WKB6 Padé & WKB6 &  difference $Re (\omega)$ & difference $Im (\omega)$ \\
\hline
$0$ & $0$ & $0.292930-0.097663 i$ & $0.292910-0.097762 i$ & $0.00678\%$ & $0.101\%$\\
$0$ & $0.2$ & $0.300251-0.095885 i$ & $0.300219-0.096021 i$ & $0.0109\%$ & $0.143\%$\\
$0$ & $0.4$ & $0.308489-0.093500 i$ & $0.308458-0.093651 i$ & $0.00978\%$ & $0.162\%$\\
$0$ & $0.6$ & $0.317903-0.090172 i$ & $0.317886-0.090311 i$ & $0.00517\%$ & $0.154\%$\\
$0$ & $0.7$ & $0.323160-0.087957 i$ & $0.323149-0.088080 i$ & $0.00341\%$ & $0.140\%$\\
$0$ & $0.75$ & $0.325945-0.086648 i$ & $0.325935-0.086764 i$ & $0.00286\%$ & $0.134\%$\\
$0.5$ & $0$ & $0.306562-0.098801 i$ & $0.306551-0.098874 i$ & $0.00368\%$ & $0.0744\%$\\
$0.5$ & $0.2$ & $0.315632-0.096253 i$ & $0.315617-0.096367 i$ & $0.00460\%$ & $0.119\%$\\
$0.5$ & $0.4$ & $0.326136-0.092551 i$ & $0.326124-0.092672 i$ & $0.00358\%$ & $0.131\%$\\
$0.5$ & $0.6$ & $0.338485-0.086674 i$ & $0.338479-0.086775 i$ & $0.00196\%$ & $0.117\%$\\
$0.5$ & $0.7$ & $0.345342-0.082222 i$ & $0.345334-0.082310 i$ & $0.00230\%$ & $0.107\%$\\
$0.5$ & $0.75$ & $0.348818-0.079417 i$ & $0.348814-0.079508 i$ & $0.00129\%$ & $0.115\%$\\
$0.9$ & $0$ & $0.352583-0.097195 i$ & $0.352625-0.097208 i$ & $0.0119\%$ & $0.0125\%$\\
$0.9$ & $0.02$ & $0.354239-0.096398 i$ & $0.354283-0.096408 i$ & $0.0122\%$ & $0.0103\%$\\
$0.9$ & $0.05$ & $0.356784-0.095088 i$ & $0.356828-0.095095 i$ & $0.0123\%$ & $0.0065\%$\\
$0.9$ & $0.1$ & $0.361170-0.092540 i$ & $0.361209-0.092543 i$ & $0.0109\%$ & $0.0034\%$\\
$0.9$ & $0.15$ & $0.365659-0.089428 i$ & $0.365691-0.089441 i$ & $0.00885\%$ & $0.0146\%$\\
$0.9$ & $0.18$ & $0.368337-0.087247 i$ & $0.368369-0.087276 i$ & $0.00874\%$ & $0.0331\%$\\
\hline
\end{tabular}
\caption{Quasinormal modes of the $\ell=1$, $n=0$ scalar field for the quantum corrected charged black hole calculated using the 6th WKB method with and without Padé approximants; $M=1$.}\label{table2}
\end{table*}

\begin{table*}
\begin{tabular}{c c c c c c}
\hline
$Q$ & $\gamma$ & WKB6 Padé & WKB6 &  difference $Re (\omega)$ & difference $Im (\omega)$ \\
\hline
$0$ & $0$ & $0.182643-0.096566 i$ & $0.182646-0.094935 i$ & $0.00155\%$ & $1.69\%$\\
$0$ & $0.2$ & $0.187696-0.094591 i$ & $0.187664-0.092974 i$ & $0.0169\%$ & $1.71\%$\\
$0$ & $0.4$ & $0.193326-0.092050 i$ & $0.193254-0.090679 i$ & $0.0372\%$ & $1.49\%$\\
$0$ & $0.6$ & $0.199779-0.088566 i$ & $0.199656-0.087533 i$ & $0.0616\%$ & $1.17\%$\\
$0$ & $0.7$ & $0.203373-0.086195 i$ & $0.203242-0.085321 i$ & $0.0644\%$ & $1.01\%$\\
$0$ & $0.75$ & $0.205248-0.084767 i$ & $0.205130-0.083963 i$ & $0.0575\%$ & $0.949\%$\\
$0.5$ & $0$ & $0.191736-0.097773 i$ & $0.191693-0.096389 i$ & $0.0222\%$ & $1.42\%$\\
$0.5$ & $0.2$ & $0.198017-0.095001 i$ & $0.197947-0.093718 i$ & $0.0351\%$ & $1.35\%$\\
$0.5$ & $0.4$ & $0.205235-0.091099 i$ & $0.205105-0.090153 i$ & $0.0633\%$ & $1.04\%$\\
$0.5$ & $0.6$ & $0.213487-0.084617 i$ & $0.213428-0.083949 i$ & $0.0273\%$ & $0.789\%$\\
$0.5$ & $0.7$ & $0.217336-0.079587 i$ & $0.217454-0.079053 i$ & $0.0544\%$ & $0.672\%$\\
$0.5$ & $0.75$ & $0.217722-0.074640 i$ & $0.219102-0.076171 i$ & $0.634\%$ & $2.05\%$\\
$0.9$ & $0$ & $0.223003-0.096157 i$ & $0.222593-0.095739 i$ & $0.184\%$ & $0.435\%$\\
$0.9$ & $0.02$ & $0.224119-0.095269 i$ & $0.223726-0.094884 i$ & $0.175\%$ & $0.405\%$\\
$0.9$ & $0.05$ & $0.225787-0.093792 i$ & $0.225456-0.093417 i$ & $0.147\%$ & $0.400\%$\\
$0.9$ & $0.1$ & $0.228465-0.090878 i$ & $0.228275-0.090450 i$ & $0.0828\%$ & $0.471\%$\\
$0.9$ & $0.15$ & $0.230806-0.087450 i$ & $0.230680-0.086968 i$ & $0.0549\%$ & $0.552\%$\\
$0.9$ & $0.18$ & $0.231963-0.085267 i$ & $0.231846-0.084767 i$ & $0.0501\%$ & $0.587\%$\\
\hline
\end{tabular}
\caption{Quasinormal modes of the $\ell=1/2$, $n=0$ Dirac field for the quantum corrected charged black hole calculated using the 6th WKB method with and without Padé approximants; $M=1$.}\label{table3}
\end{table*}

\begin{table*}
\begin{tabular}{c c c c c c}
\hline
$Q$ & $\gamma$ & WKB6 Padé & WKB6 &  difference $Re (\omega)$ & difference $Im (\omega)$ \\
\hline
$0$ & $0$ & $0.380041-0.096408 i$ & $0.380068-0.096366 i$ & $0.00718\%$ & $0.0435\%$\\
$0$ & $0.2$ & $0.389393-0.094682 i$ & $0.389434-0.094619 i$ & $0.0107\%$ & $0.0662\%$\\
$0$ & $0.4$ & $0.400077-0.092364 i$ & $0.400133-0.092281 i$ & $0.0139\%$ & $0.0901\%$\\
$0$ & $0.6$ & $0.412551-0.089087 i$ & $0.412615-0.088987 i$ & $0.0154\%$ & $0.112\%$\\
$0$ & $0.7$ & $0.419672-0.086872 i$ & $0.419736-0.086767 i$ & $0.0153\%$ & $0.121\%$\\
$0$ & $0.75$ & $0.423503-0.085554 i$ & $0.423566-0.085447 i$ & $0.0149\%$ & $0.125\%$\\
$0.5$ & $0$ & $0.397890-0.097601 i$ & $0.397918-0.097551 i$ & $0.00710\%$ & $0.0515\%$\\
$0.5$ & $0.2$ & $0.409617-0.095127 i$ & $0.409656-0.095064 i$ & $0.00948\%$ & $0.0655\%$\\
$0.5$ & $0.4$ & $0.423497-0.091496 i$ & $0.423547-0.091419 i$ & $0.0117\%$ & $0.0846\%$\\
$0.5$ & $0.6$ & $0.440487-0.085587 i$ & $0.440538-0.085497 i$ & $0.0117\%$ & $0.104\%$\\
$0.5$ & $0.7$ & $0.450537-0.080903 i$ & $0.450585-0.080814 i$ & $0.0105\%$ & $0.110\%$\\
$0.5$ & $0.75$ & $0.455964-0.077799 i$ & $0.456008-0.077713 i$ & $0.00974\%$ & $0.111\%$\\
$0.9$ & $0$ & $0.458819-0.096251 i$ & $0.458840-0.096215 i$ & $0.00451\%$ & $0.0375\%$\\
$0.9$ & $0.02$ & $0.461076-0.095465 i$ & $0.461097-0.095428 i$ & $0.00469\%$ & $0.0387\%$\\
$0.9$ & $0.05$ & $0.464581-0.094167 i$ & $0.464604-0.094128 i$ & $0.00494\%$ & $0.0413\%$\\
$0.9$ & $0.1$ & $0.470765-0.091609 i$ & $0.470789-0.091566 i$ & $0.00511\%$ & $0.0469\%$\\
$0.9$ & $0.15$ & $0.477378-0.088396 i$ & $0.477402-0.088350 i$ & $0.00499\%$ & $0.0513\%$\\
$0.9$ & $0.18$ & $0.481529-0.086053 i$ & $0.481553-0.086007 i$ & $0.00510\%$ & $0.0530\%$\\
\hline
\end{tabular}
\caption{Quasinormal modes of the $\ell=3/2$, $n=0$ Dirac field for the quantum corrected charged black hole calculated using the 6th WKB method with and without Padé approximants; $M=1$.}\label{table4}
\end{table*}

\begin{table*}
\begin{tabular}{c c c c c}
\hline
$\gamma $ & time-domain & WKB6 Padé &  error $Re (\omega)$ & rel. error $Im (\omega)$  \\
\hline
$0$ & $0.115957-0.106015 i$ & $0.116107-0.105495 i$ & $0.129\%$ & $0.491\%$\\
$0.1$ & $0.117937-0.104554 i$ & $0.118101-0.103991 i$ & $0.140\%$ & $0.538\%$\\
$0.2$ & $0.119924-0.102806 i$ & $0.120011-0.102341 i$ & $0.0727\%$ & $0.453\%$\\
$0.3$ & $0.121869-0.100692 i$ & $0.121699-0.100549 i$ & $0.140\%$ & $0.142\%$\\
$0.4$ & $0.123673-0.098096 i$ & $0.123373-0.098171 i$ & $0.243\%$ & $0.0767\%$\\
$0.5$ & $0.125112-0.094870 i$ & $0.125308-0.095008 i$ & $0.157\%$ & $0.145\%$\\
$0.6$ & $0.125597-0.090946 i$ & $0.126313-0.090288 i$ & $0.570\%$ & $0.724\%$\\
$0.7$ & $0.123817-0.087671 i$ & $0.124233-0.087455 i$ & $0.336\%$ & $0.247\%$\\
$0.75$ & $0.123276-0.087287 i$ & $0.123657-0.086991 i$ & $0.309\%$ & $0.339\%$\\
\hline
\end{tabular}
\caption{Comparison of the quasinormal frequencies obtained by the time-domain integration and the 6th order WKB approach with Padé approximants for $s=\ell=0$ ($M=1$, $Q=0.5$).}\label{check1}
\end{table*}

\begin{table*}
\begin{tabular}{c c c c c}
\hline
$\gamma $ & time-domain & WKB6 Padé & error $Re (\omega)$ & rel. error $Im (\omega)$  \\
\hline
$0$ & $0.306577-0.098796 i$ & $0.306562-0.098801 i$ & $0.00480\%$ & $0.0053\%$\\
$0.1$ & $0.310960-0.097625 i$ & $0.310941-0.097634 i$ & $0.00621\%$ & $0.0090\%$\\
$0.2$ & $0.315654-0.096237 i$ & $0.315632-0.096253 i$ & $0.00703\%$ & $0.0165\%$\\
$0.3$ & $0.320704-0.094569 i$ & $0.320681-0.094590 i$ & $0.00714\%$ & $0.0217\%$\\
$0.4$ & $0.326163-0.092526 i$ & $0.326136-0.092551 i$ & $0.00842\%$ & $0.0272\%$\\
$0.5$ & $0.332089-0.089962 i$ & $0.332059-0.089992 i$ & $0.00903\%$ & $0.0331\%$\\
$0.6$ & $0.338516-0.086642 i$ & $0.338485-0.086674 i$ & $0.00914\%$ & $0.0366\%$\\
$0.7$ & $0.345364-0.082186 i$ & $0.345342-0.082222 i$ & $0.00623\%$ & $0.0432\%$\\
$0.75$ & $0.348835-0.079390 i$ & $0.348818-0.079417 i$ & $0.00478\%$ & $0.0347\%$\\
\hline
\end{tabular}
\caption{Comparison of the quasinormal frequencies obtained by the time-domain integration and the 6th order WKB approach with Padé approximants for $s=0$, $\ell=1$ ($M=1$, $Q=0.5$).}\label{check2}
\end{table*}

\begin{table*}
\begin{tabular}{c c c c c}
\hline
$\gamma $ & time-domain & WKB6 Padé & error $Re (\omega)$ & rel. error $Im (\omega)$ \\
\hline
$0$ & $0.192194-0.098007 i$ & $0.191736-0.097773 i$ & $0.238\%$ & $0.238\%$\\
$0.1$ & $0.195248-0.096726 i$ & $0.194783-0.096487 i$ & $0.238\%$ & $0.247\%$\\
$0.2$ & $0.198519-0.095216 i$ & $0.198017-0.095001 i$ & $0.253\%$ & $0.226\%$\\
$0.3$ & $0.202036-0.093409 i$ & $0.201482-0.093247 i$ & $0.274\%$ & $0.174\%$\\
$0.4$ & $0.205817-0.091194 i$ & $0.205235-0.091099 i$ & $0.283\%$ & $0.105\%$\\
$0.5$ & $0.209856-0.088393 i$ & $0.209279-0.088332 i$ & $0.275\%$ & $0.0683\%$\\
$0.6$ & $0.214064-0.084711 i$ & $0.213487-0.084617 i$ & $0.270\%$ & $0.111\%$\\
$0.7$ & $0.218045-0.079771 i$ & $0.217336-0.079587 i$ & $0.325\%$ & $0.230\%$\\
$0.75$ & $0.219675-0.076860 i$ & $0.217722-0.074640 i$ & $0.889\%$ & $2.89\%$\\
\hline
\end{tabular}
\caption{Comparison of the quasinormal frequencies obtained by the time-domain integration and the 6th order WKB approach with Padé approximants for $s=1/2$, $\ell=1/2$ ($M=1$, $Q=0.5$).}\label{check3}
\end{table*}

\section{Quasinormal modes}\label{sec:QNM}

Quasinormal modes of a scalar field for the quantum corrected charged black hole has been recently considered in \cite{Javed:2024hzs}. The results for $\ell=1$ are summarized in table 3 and figs. 12 and 13. No data for other values $\ell$, including the fundamental mode $\ell=n=0$ were presented in  \cite{Javed:2024hzs}. Therefore, for the purpose of comparison, we will refer back to table 3. We start from the Schwarzschild limit, corresponding to $\gamma =Q=0$  and $\ell=1$ for which the author of \cite{Javed:2024hzs} obtains $\omega = 0.19245 - 0.096225 i$. The precise value of this frequency calculated by the Leaver method \cite{Leaver:1985ax} is well-known to be $\omega = 0.292936 - 0.09766 i$, which means an enormous numerical error leading to about 1.5 time smaller real oscillation frequency. Notice that the effect due to non-zero $\gamma$ reaches only several percents and about one order smaller than the error. At the same time, in our case, as shown in the table \ref{table2} the 6th order WKB formula with the Padé approximants gives $\omega =0.292930 - 0.097663 i$, which almost completely coincides with the precise Leaver result. Similarly, we can see that our data for $\gamma = 0.4$, $Q=0$ gives $\omega = 0.326163 - 0.092526 i$ by time-domain integration and  $\omega = 0.326136 - 0.092551 i$ by the WKB method, which is again in excellent agreement between each other and differs a lot from the value $\omega =  0.202224 - 0.0923522 i$ obtained in \cite{Javed:2024hzs}.

The explanation of the failure of the calculations suggested in \cite{Javed:2024hzs} is the usage of the eikonal formula which is accurate only at $\ell=\infty$ but has a very big error at $\ell=1$. Then, it is natural that no fundamental mode $\ell=n=0$ was presented in \cite{Javed:2024hzs}, because the eikonal formula would be even worse for this case.
On the contrary, usage of the 6th order WKB formula (instead of the first order one) and farther application of the  Padé approximants provides sufficiently accurate results which are in good agreement with the time-domain integration.

Examples of the time-domain profile for the $\ell =0$ and $\ell=1$ scalar field perturbations are given in figures \ref{fig:timedomain} and  \ref{fig:timedomain2} respectively. Time-domain profiles for Dirac perturbations are similar. 
From tables \ref{check1}- \ref{check3} we can see that the 6th order WKB method with the Padé approximants are in a very good concordance with the time-domain integration, while the usual 6th order WKB formula is slightly less accurate for $\ell >0$, but may reach several percents for $\ell=0$ scalar perturbations. Taking the time-domain integration data as accurate for the lowest multipole numbers we can see that the relative error of the 6th order WKB method with the Padé approximants usually does not exceed a small fraction of one percent, except a single case of $\ell=1/2$ Dirac perturbations of the near extremely charged black hole for which the relative error exceeds two percents. In that case, we should rely on time-domain integration to a greater extent than the WKB method, because, the latter converges only asymptotically and the convergence is not guaranteed in each consequent order.

From tables \ref{table1}-\ref{check3} we see that when the quantum parameter $\gamma$ is tuned on, $Re \omega$ is noticeably increased, while $Im \omega$ decreases.
This means that the quality factor, proportional to the ratio of the real oscillation frequency to the damping rate, is considerably increased, and, consequently, the quantum corrected black hole is much better oscillator than the classical one.  

\section{Eikonal formula}

As a rule quasinormal modes can be calculated only numerically. However, in the eikonal limit the analytical expressions can be obtained. While this limit is discussed in \cite{Javed:2024hzs}, no derivation of the explicit formula in terms of the parameters of the system, such as mass $M$, charge $Q$ and the coupling $\gamma$ are done. 
Here we will fill this gap.

Perturbations of spherically symmetric black holes can be simplified to the wave-like equation with the effective potential approximated as follows:
\begin{equation}\label{potential-multipole}
V(r_*)=\kappa^2\left(H(r_*)+\Order{\kappa^{-1}}\right),
\end{equation}
where $\kappa\equiv\ell+\frac{1}{2}$ and $\ell=s,s+1,s+2,\ldots$ represents the positive multipole number. Its minimum value equals the spin of the field $s$. Following the conventions of \cite{Konoplya:2023moy}, we expand in powers of $\kappa^{-1}$.

The function $H(r_*)$ exhibits a single peak. Thus, the position of the potential's maximum (\ref{potential-multipole}) can be approximated as:
\begin{equation}\label{rmax}
  r_{\max }=r_0+r_1\kappa^{-1}+r_2\kappa^{-2}+\ldots.
\end{equation}

Substituting (\ref{rmax}) into the first order WKB formula
\begin{eqnarray}
\omega&=&\sqrt{V_0-\imo \K\sqrt{-2V_2}},
\end{eqnarray}
and expanding in $\kappa^{-1}$, we obtain,
\begin{eqnarray}\label{eikonal-formulas}
\omega=\Omega\kappa-\imo\lambda\K+\Order{\kappa^{-1}}.
\end{eqnarray}
The above relation is a good approximation in the regime $\kappa\gg\K$.
Finally, expanding also in powers of $Q$, we obtain:
\begin{widetext}
$$ r_{\max} = \left(\left(3 M-\frac{2 Q^2}{3 M}+O\left(Q^3\right)\right)+\gamma  \left(-\frac{5}{9 M}-\frac{7 Q^2}{27 M^3}+O\left(Q^3\right)\right)+O\left(\gamma^2\right)\right)+ $$
\begin{equation}
\frac{O\left(Q^3\right)+O\left(Q^3\right) \gamma +O\left(\gamma ^2\right)}{\kappa}+O\left(\left(\frac{1}{\kappa }\right)^2\right)
\end{equation}
$$ \omega = \kappa  \left(\left(\frac{1}{3 \sqrt{3} M}+\frac{Q^2}{18
   \sqrt{3} M^3}+O\left(Q^3\right)\right)+\left(\frac{1}{27
   \sqrt{3} M^3}+\frac{7 Q^2}{243 \sqrt{3}
   M^5}+O\left(Q^3\right)\right) \gamma +O\left(\gamma
   ^2\right)\right)+ $$
\begin{equation}
\left(\left(-\frac{i \K}{3 \sqrt{3}
   M}-\frac{i Q^2 \K}{54 \sqrt{3}
   M^3}+O\left(Q^3\right)\right)+\gamma  \left(\frac{2 i
   \K}{81 \sqrt{3} M^3}+\frac{7 i \K Q^2}{243 \sqrt{3}
   M^5}+O\left(Q^3\right)\right)+O\left(\gamma
   ^2\right)\right)+O\left(\frac{1}{\kappa}\right).
\end{equation}
\end{widetext}

The correspondence suggested in \cite{Cardoso:2008bp} claims that particular parameters (Lyapunov exponent and angular velocity) of the unstable circular null geodesics around static, spherically symmetric black holes are linked to the quasinormal modes of the black hole in the high frequency limit  $\ell \gg n$ regime: 
\begin{equation}\label{QNM}
\omega_n=\Omega\ell-\imo(n+1/2)|\lambda|, \quad \ell \gg n.
\end{equation}
Here $\Omega$ is the angular velocity at the unstable circular null geodesics, and $\lambda$ is the Lyapunov exponent.
This correspondence holds for a great number of cases and there is an extensive literature where the analytic formulas for eikonal quasinormal modes are deduced see, for instance \cite{Bolokhov:2023bwm,Abdalla:2005hu,Zhidenko:2008fp,Konoplya:2018ala,doi:10.1142/S0217751X24500246,Dolan:2010wr,Konoplya:2005sy,Hod:2009td,Jusufi:2020dhz}). However, as was shown in \cite{Konoplya:2017wot,Konoplya:2020bxa,Konoplya:2019hml,Konoplya:2022gjp,Bolokhov:2023dxq} it breaks down in various cases, because the correspondence is based on the similarity between characteristics of the null geodesics and the first order WKB formula. When the WKB is inadequate or insufficient for description the spectrum, the correspondence breaks. Thus, it is useful to check it for each case under consideration.
Here we can see that the correspondence is indeed verified for scalar and Dirac perturbations.

\section{Conclusions}\label{sec:conclusions}

In the present paper we have calculated dominant quasinormal modes for scalar and Dirac/neutrino perturbations of quantum corrected charged black hole constructed in Asymptotically Safe Gravity. We show that the previous publication \cite{Javed:2024hzs} on scalar quasinormal modes of this black hole contain a huge numerical error, exceeding by one order the total effect.  
In contrast, our work demonstrates excellent agreement between time-domain integration and the 6th-order WKB method with Padé approximants. We further contribute by deriving an explicit analytic formula for quasinormal modes in the eikonal regime.

\begin{acknowledgments}
The author would like to thank R. A. Konoplya for his useful discussions and kind help.  
\end{acknowledgments}

\bibliographystyle{unsrt}
\bibliography{bibliography}
\end{document}